\documentclass[aip,sd,amsmath,amssymb,reprint]{revtex4-1}

\usepackage{graphicx}
\usepackage{dcolumn}
\usepackage{bm}
\usepackage[utf8]{inputenc}
\usepackage[T1]{fontenc}
\usepackage{mathptmx}
\DeclareUnicodeCharacter{2212}{-}
\begin{document}

\title{Charging of a Single InAs QD with Electrically-Injected Holes using a Lateral Electric Field}

\author{Xiangyu Ma$^{1}$, Yuejing Wang$^{2}$, Joshua Zide $^{2}$, Matthew Doty}
\affiliation{Dept. of Physics and Astronomy, University of Delaware, Newark, DE 19716}
\affiliation{Material Science and Engineering Department, University of Delaware, Newark, DE 19716}
\date{\today}

\begin{abstract}
InAs/GaAs quantum dots (QDs) and quantum dot molecules (QDMs) are self-assembled semiconductor nanostructures that can trap a single electron or hole with well-defined spin projections. QDs and QDMs have excellent optical properties and have long been of interest for incorporation into quantum optoelectronic devices ranging from single photon sources to multi-bit quantum computers. The properties of single QDs, or carriers confined within those QDs, can be tuned by external electric fields, which provides an important tool for the development of scalable and tunable devices. Deterministic charging of a QD with a single electron or hole is an important tool for quantum devices and is well-established under the application of growth-direction electric fields in a diode structure. Here, we report a new charging mechanism for a single QD in a 3-electrode device that does not contain a vertical diode and can be used to control electric field profiles in two dimensions. We fabricate the device with E-beam lithography and characterize photoluminescence from single QDs under different bias configurations. Using a combination of experimental data and COMSOL band structure calculations, we explain how the charging originates in the electrical-injection of holes induced by lateral electric fields. We discuss the applications for this device and the potential for full 2-D electric field control of a single QD and QDM. 
\end{abstract}


\maketitle

\section{\label{sec1}Introduction:\protect}
Single self-assembled InAs/GaAs quantum dots (QDs) and quantum dot molecules (QDMs) grown by Molecular Beam Epitaxy (MBE) have excellent optical properties for quantum applications such as single photon generation,\cite{ding2016demand,somaschi2016near,hanschke2018quantum} spin-photon coupling,\cite{davanco2017heterogeneous,dory2016complete, hennessy2007quantum} and spin manipulation.\cite{warburton2013single, doty2010opportunities} A scalable quantum optical platform requires a large number of identical quantum emitters, spin units, and so forth. However, the random nucleation of the Stranski-Krastanov growth process typically leads to QDs and QDMs with different sizes, shapes and geometries.\cite{gammon1996fine, yamaguchi2000stranski, doty2010hole} External electric fields are commonly used to fine-tune the optical properties of a QD to overcome these variations and to control the charge occupancy of the QD. Specifically, a QD is usually charged with electrons or holes by embedding the QD in a vertical (growth direction) diode structure and applying a vertical electric field that tunes the QD energy levels relative to the Fermi level set by the doping.\cite{warburton2013single,zhou2011spectroscopic} This same vertical electric field is often used to tune the QD emission wavelength via the Stark Shift or, for QDMs, via indirect optical transitions.\cite{stinaff2006optical, bracker2006engineering, doty2006electrically, liu2011situ} Lateral electric fields have also been used to fine tune the emission energies, fine structure splittings (FSS) \cite{gerardot2007manipulating,reimer2008prepositioned,kowalik2005influence}, and spin lifetime \cite{moody2016electronic}, but these lateral fields have been applied in the absence of growth direction (vertical) fields and have typically relied on either neutral excitons or spontaneous optical charging of the QDs. Simultaneously applying both vertical and lateral electric fields to create a 2-D electric field profile in the vicinity of a single QD could achieve simultaneous control over charging, emission wavelength, and FSS. Such a 2-D electric field profile could even be used to controllably induce novel spin properties for qubit applications.\cite{zhou2014design, ma2016hole, doty2010hole,economou2012scalable}

In this paper, we demonstrate charging of a single QD with electrically-injected holes whose tunneling into the QDs is induced by lateral electric fields. This device and result represents an important first step toward controlled generation of arbitrary 2-D electric fields at the location of a single QD and reveals important new physical phenomena that will need to be managed for device applications. We first introduce a 3-electrode device structure and the device fabrication process. We then report micro-photoluminescence (micro-PL) measurements of a single QD when we apply different biases to this device. We use the charging observed in the micro-PL data and band structure simulation results from COMSOL to establish that the charging of a single QD arises from lateral electric fields. 

\begin{figure}[h]
  \centering
  \includegraphics[width=8cm]{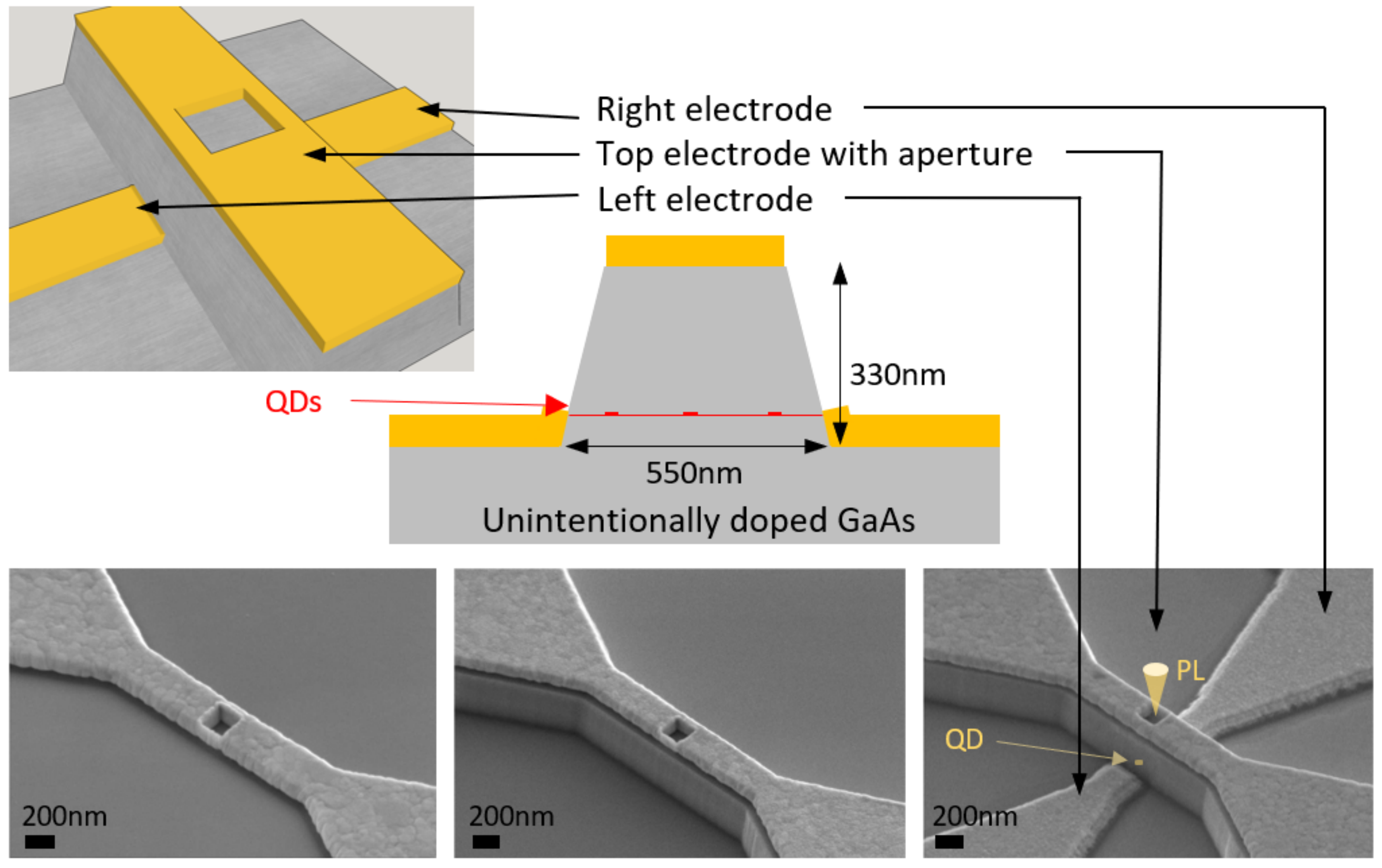}
\caption{Design and key fabrication step of 3-electrode device. SEM images of (left) Top electrode and aperture; (middle) Mesa (right) Lateral electrodes}
\label{DesignFab}
\end{figure}

\section{\label{sec2}Device Design and Fabrication}

\begin{figure}[ht]
  \centering
  \includegraphics[width=8cm]{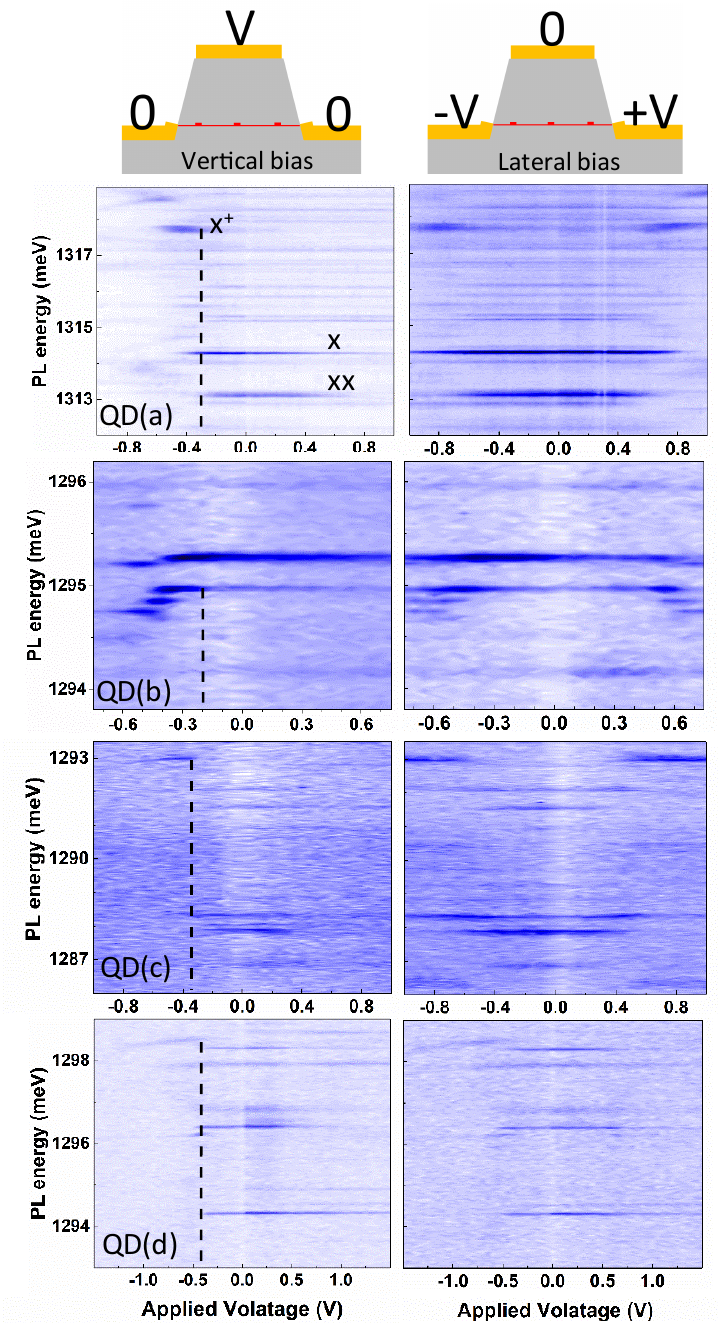}
\caption{Micro-photoluminescence of four different QDs from 4 different apertures on the 3-electrode device. The left panel shows the PL intensity map under a vertical bias, and the right panel shows the PL intensity under a lateral bias. The lateral axis represents the applied voltage V in each case, and the vertical axis represents the PL energy observed from a single aperture. }
\label{PL}
\end{figure}

We grow our QDs using MBE on an intrinsic GaAs (001) substrate. We first grow 500nm unintentionally-doped GaAs on top of the substrate. We then deposit approximately 1.7ML of InAs for QD nucleation and truncate the height of the dots with a 2.7nm GaAs partial cap and subsequent flush of the exposed In at elevated tempearture. We then grow another 280nm unintentionally-doped GaAs on top of the QD for protection. We stop the substrate rotation briefly during QD growth to form a QD density gradient, allowing us to locate a region of the wafer with relatively low QD density such that, on average, only 1-3 QDs lie within each device. We performed Hall effect measurements to determine that the unintentionally-doped GaAs is slightly p-type with a bulk carrier concentration of $5\times10^{16}/cm^3$.

The active part of our device consists of three electrodes patterned around an unintentionally-doped GaAs mesa that contains InAs QDs, as shown in Figure \ref{DesignFab}. The mesa has a trapezoidal shape, and the side walls have a slope of 15$^\circ$. The InAs QDs are embedded near the bottom of the GaAs mesa. Two lateral electrodes are patterned on the side of the mesa, with a sidewall coverage slightly above the height of the QDs. The top electrode has a 250nm $\times$ 250nm wide aperture, allowing optical access to the QD from the top of the wafer. We use E-beam lithography to fabricate the device because the fabrication requires a layer to layer alignment error smaller than 10nm. We first fabricate the top electrode (20nm Ti + 100nm Au) using a lift-off process with PMMA/MMA bi-layer resist and e-beam metal deposition. We then pattern the aperture on the top electrode and use Ion Milling to etch away the Au. The Ion Mill etching recipe has a high selectivity between Au and Ti so that the etch can stop at the Ti layer without damaging the underlying GaAs. The 20nm Ti is semi-transparent in the NIR wavelength range. We then etch the mesa using $BCl_3$ Inductively Coupled Plasma (ICP) etching, creating the trapezoidal mesa shape with the angled sidewalls. To fabricate the lateral electrodes we use a bi-layer resist, angled e-beam metal deposition, and final lift-off process. In order to guarantee that the lateral electrodes cover the mesa sidewalls up to a point just slightly higher than the height of the QDs, we calculate the geometry of the resist structure after characterizing it using SEM, then calculate the appropriate metal deposition angle. SEM images of several key steps during the fabrication are shown in Figure \ref{DesignFab}. To maximize our chances of observing a single QD in an aperture, we fabricate 180 individual apertures on a 1cm*1cm die, with each one identical to the structure shown in Figure \ref{DesignFab}. These electrodes are connected to 500um*500um bonding pads for wire-bonding. 

\section{\label{sec3}Device Characterization}
\subsection{Micro-photoluminescence}

After wire-bonding, we mount the sample in an ARS-DMX20 cryostat to study PL at 8K. We use a LED light to locate each aperture and a 5uW 780nm CW laser to excite the quantum dots. We focus the laser light with a NIR objective with a spot size close to the diffraction limit so that we will only excite QDs from a single aperture. We collect the PL through two OD5 900nm long pass filters and measure the energy of the emitted photons with a NIR spectrometer equipped with a liquid-nitrogen-cooled CCD. We use a Keithley 2230 multi-channel voltage source to apply voltages to the device. 

We first survey all the apertures to find the ones that contain a single QD, then apply voltages for detailed study of the selected apertures. Figure \ref{PL} shows the PL intensity graph from 4 different apertures (QDs) under different electric field conditions. All PL is collected with the device current smaller than 500uA, close to our current detection limit. For all data presented in the left column, we ground the two lateral electrodes and apply the voltage V to the top electrode. We name this bias configuration the vertical bias. Under these conditions, we observe an asymmetric PL charging map centered at 0V in all 4 QDs presented here. We assign the charge states based on excitation power dependence (data not shown) and literature reports of the typical energy shifts between various charge configurations.\cite{zhou2011spectroscopic,stinaff2006optical,ware2005polarized} For example, in QD (a), the positively charged state ($X^+$) with energy 1317.8meV becomes visible at -0.3V, as indicated by the dashed line. The neutral exciton state ($X$) with energy 1314.3meV and the neutral biexciton state ($XX$) with energy 1313.1meV disappear at the same voltage. Although the neutral exciton states disappear when we apply a sufficiently large positive voltage to the top electrode, we do not see a positively charged state appear. We therefore attribute the loss of signal from the neutral exciton states to electric fields that allow at least one carrier to escape the QD before recombination. We observe similar photoluminescence features in QD (b), (c), and (d). The precise energy differences between the neutral exciton state and the positively charged state vary between different dots. This is likely due to the different electron-hole interaction strengths that result from variations in QD size and composition.

For the data presented in the right column, we apply equal and opposite voltage to the two lateral electrodes while grounding the top electrode. We name this bias configuration the lateral bias. We apply the same magnitude of voltages as the vertical bias. In this case, we observe a charging sequence in each QD exactly analogous to what is observed in the left column. The difference is the symmetry. For example: in QD (a) the positively charged exciton emerges symmetrically at both V=0.6V and V=-0.6V. The same symmetric charging PL map can be seen in QD (b) and (c). This symmetric charging pattern is representative of the majority of the apertures that we surveyed. Very occasionally, we observe asymmetric charging features similar to those obtained under the vertical bias, as shown in QD (d). The faint charged line around 1288.5meV shows up only on one side of the lateral PL bias map, in contrast to the symmetric charging pattern observed in apertures a, b, and c.

To further understand this charging behavior, we conduct wavelength-dependent PL measurements of these QDs. As shown in the Supplementary material, the PL suddenly quenches when the excitation wavelength increases from 820nm to 830nm, indicating a lack of carriers available for excitation when the laser wavelength is longer than 830nm. This wavelength corresponds to the energy of the p-type defects formed during growth. These defects push the Fermi level close to the valence band-edge, and large band-bending can be formed near the Schottky contact. At 8K, holes in these defects are likely to freeze or relax to the wetting layer valence band because their energy levels are close. We suspect that holes trapped in the wetting layer are the reservoir for the electrically-injected holes observed in the PL spectra. 

\begin{figure}[ht]
  \centering
  \includegraphics[width=8cm]{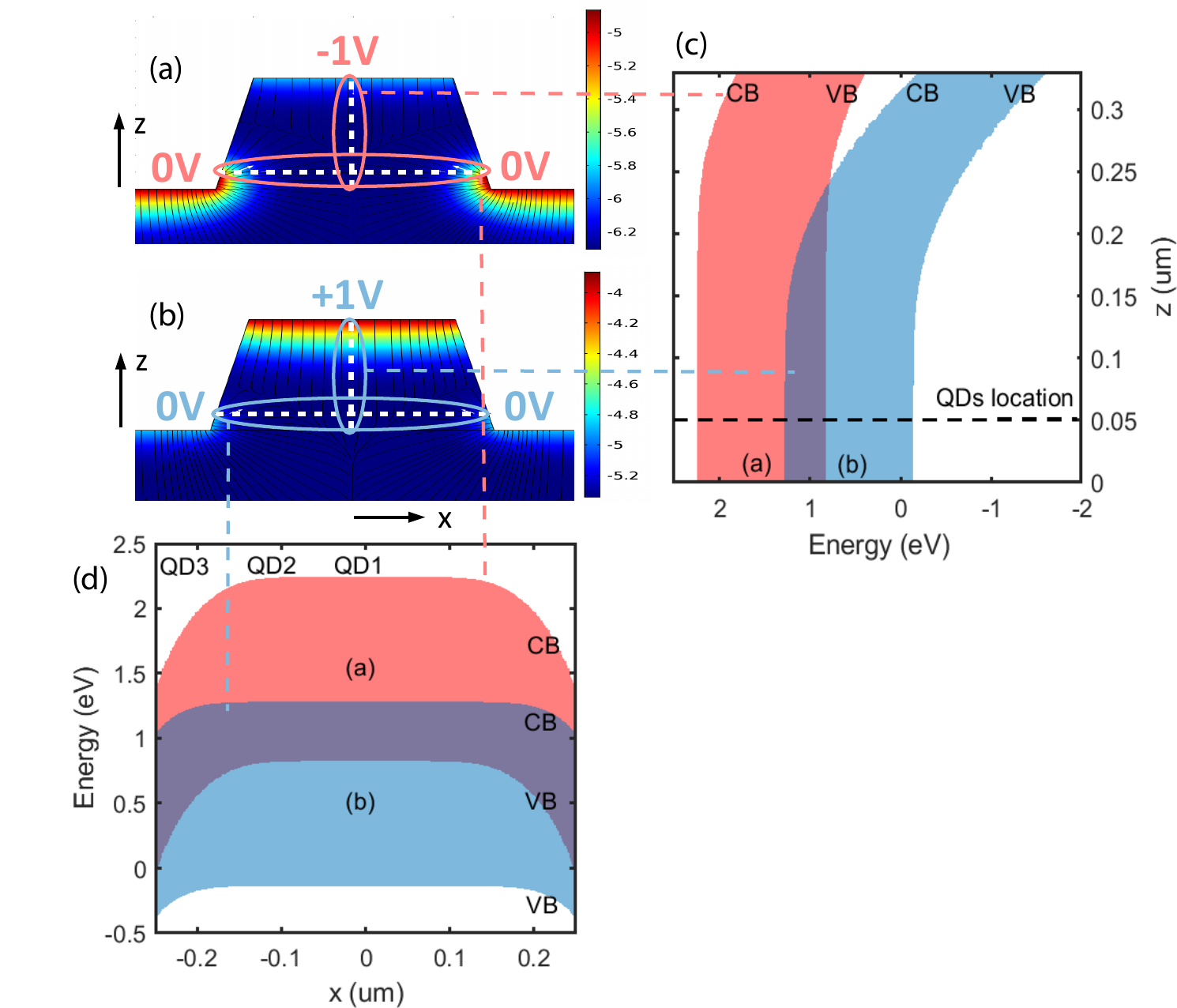}
\caption{The electric potential and band diagrams of the 3-electrode device under (a) negative vertical bias and (b) positive vertical bias. The vertical white line in (a) and (b) follows the center of the mesa. The horizontal white line represents the location of the QDs. The curves in (c) and (d) show the 1-D band structure along the vertical and horizontal cut-lines, respectively. }
\label{COMSOL1}
\end{figure}

\subsection{COMSOL simulation}

\begin{figure}[ht]
  \centering
  \includegraphics[width=8cm]{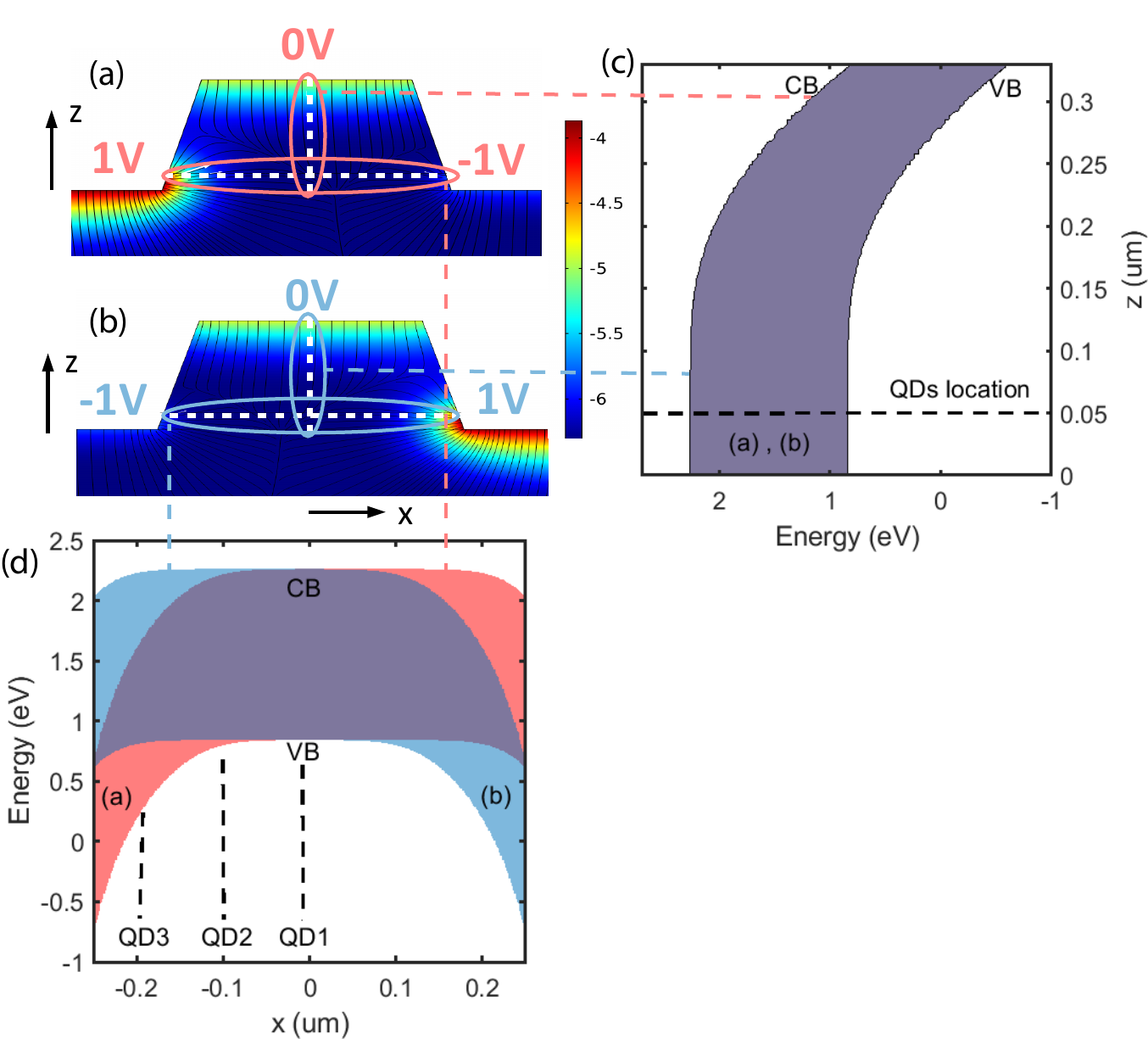}
\caption{The electric potential and band diagrams of the 3-electrode device under (a) negative lateral bias and (b) positive lateral bias. The vertical white line in (a) and (b) follows the center of the mesa. The horizontal white line represents the location of the QDs. The curves in (c) and (d) show the 1-D band structure along the vertical and horizontal cut-lines, respectively. }
\label{COMSOL2}
\end{figure}

To explain the observed charging in the absence of a traditional n- or p-doped layer or a conventional diode, we calculate the band structure of this device using COMSOL simulations with the semiconductor model. We build the device model with a geometry mimicking the device structures in the SEM picture. For non-polar semiconductors like GaAs, we use 0.8V as the Schottky barrier height for the Ti/Au metal contact.\cite{monch1999barrier, altuntacs2009electrical} Because the exact Fermi level of the device is unknown at low temperature, we used the Fermi level derived from the doping level in GaAs, which qualitatively agrees with the wavelength-dependent PL data. We calculate the electric potential and the band-diagram in both the vertical bias and the lateral bias configurations, as shown in Figure \ref{COMSOL1} and Figure \ref{COMSOL2}. 

The colored surface map in Figures \ref{COMSOL1} (a) and (b) shows the electric potential of the 3-electrode device when we apply a vertical bias. When we apply -1V to the top electrode, the bias forms a large potential gradient near the two lateral electrodes, while the rest of the mesa's potential remains relatively flat. When we apply +1V to the top electrode, the bias forms a large potential gradient near the top of the mesa. More direct views of the band structure along the two white dashed lines are shown in Figures \ref{COMSOL1} (c) and (d). When we apply -1V to the top electrode, there is a large lateral electric field near the edge of the mesa, while the vertical band remains flat, as shown by the red data in Figures \ref{COMSOL1} (c) and (d), respectively. In contrast, when we apply +1V to the top electrode the bias forms little lateral electric field but generates a large vertical electric field along the vertical dashed line, as shown by the blue data in Figures \ref{COMSOL1} (c) and (d). 

The experimental data show that charging only happens when we apply a negative voltage to the top electrode, indicating that holes are electrically injected into the QDs by the lateral electric fields formed on both sides of the mesa. We do not observe hole charging when we apply a positive voltage to the top electrode because the lateral electric field is small. This is why the PL diagram under the vertical bias is asymmetric in all observed cases, represented by Figure \ref{PL}.

The colored surface map in Figures \ref{COMSOL2} (a) and (b) shows the electric potential of the 3-electrode device when we apply lateral biases. When we apply 1V on the left electrode and -1V on the right electrode, a large potential gradient is formed near the left electrode, as shown in Figure \ref{COMSOL2} (a). The electric potential diagram is flipped if we apply -1V on the left electrode and +1V on the right electrode, i.e. we observe a large potential gradient near the right electrode as shown in Figure \ref{COMSOL2} (b). Again the band structures along the two dashed lines are shown in Figures \ref{COMSOL2} (c) and (d). In the vertical direction (Figure \ref{COMSOL2} (c)), both biases generate an electric field near the top of the mesa. In the lateral direction (Figure \ref{COMSOL2} (d)), both biases can result in a large lateral electric field on the side of the mesa that has the positive bias. 

For QDs that are near the center of the mesa, e.g. the locations denoted by QD1 and QD2 in Figure \ref{COMSOL2}(d), holes can be injected into QDs with either a positive lateral bias or a negative lateral bias because of the large lateral electric field formed near each lateral electrode when it is positively charged. This is likely the case for QD (a), (b) and (c) in Figure \ref{PL}. For QDs that reside near the edge of the mesa, such as the location denoted by QD3 in Figure \ref{COMSOL2}(d), holes are only likely to be injected into the QD when there's a large electric field on that side of the mesa. This is likely the case for QD (d) in Figure \ref{PL}. 

We note that the COMSOL simulations, which are consistent with the experimental data, indicate that this devices has generated a 2-D electric field. A very clear indication of the 2-D electric field control would be observation of a Stark shift from both the vertical and lateral electric fields. Although we do observe some Stark shift at high voltages, the overall magnitude is very small. This indicates that the magnitude of the electric field at the QD is small. To increase the magnitude of the electric field in the future, we can reduce the p-type defect density in the intrinsic GaAs. This would increase the depletion width and reduce the band-bending near the Schottky contact. We can also introduce a $\delta$ doping layer near the QD height to charge a single QD with an electron or a hole. 

\section{\label{sec5}Conclusion}
We report design and fabrication of a 3-electrode device that can apply 2-D electric fields to single QDs embedded in an unintentionally-doped GaAs matrix. We characterized single QD PL under different bias configurations and observed an asymmetric charging pattern under a vertical bias and a symmetric charging pattern under a lateral bias. Combining the experimental data with electric field simulations, we deduce that these charging events likely originate from electrically-injected holes whose tunneling into the QDs is induced by lateral electric fields. This device and the observed results represent a first step toward generation of arbitrary 2-D electric fields applied to single QDs or QDMs and, to the best of our knowledge, the first observation of deterministic charging of single QDs arising from lateral electric field components. 

\section{\label{sec6}Acknowledgement}
We acknowledge financial support from the NSF (DMR-1505574). We also acknowledge support from Allan Bracker and Michael Yakes of the Naval Research Laboratory, who helped us to grow test samples and provided valuable suggestions. Finally, we acknowledge support from Iulian Codreanu, Kevin Lister, Paul Horng and Scott McCracken of the UD Nanofabrication Facility, who helped us to develop the fabrication recipes employed.

\section{\label{refe}References}

\appendix
\section{Wavelength-dependent PL}
\begin{figure}[h]
  \centering
  \includegraphics[width=8cm]{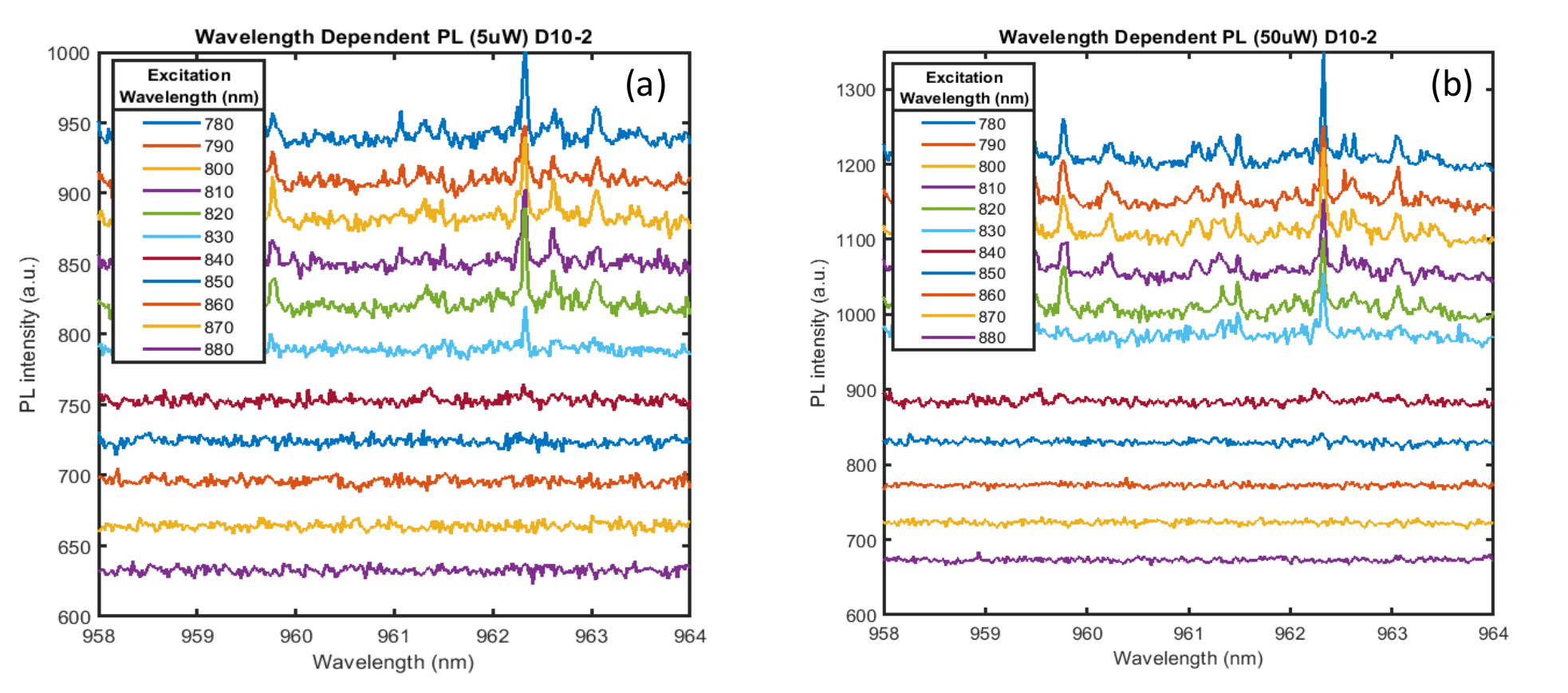}
\caption{Wavelength dependent study of a single QD PL. The same wavelength threshold for QD PL is found in measurement with 5uW (a) and 50uW (b) excitation powers. }
\label{Wavelengthd}
\end{figure}

The PL of QD(c) as a function of excitation wavelength is shown in Figure \ref{Wavelengthd}. The PL intensity begins to drop significantly when the excitation wavelength increases from 820nm to 830nm and reaches zero for excitation wavelengths longer than 830 nm. This indicates that carriers involved in PL emission are generated by excitation across the GaAs bandgap, with weaker generation through near-band-gap impurity states that typically emit at around 830 nm. This excitation wavelength threshold does not change with different excitation powers, as compared in Figure \ref{Wavelengthd}(a) and (b).

\end{document}